\documentclass[aps,prl,twocolumn,showpacs,amsmath,amssymb,floatfix,superscriptaddress,longbibliography]{revtex4-2}
\usepackage{graphicx}
\usepackage{dcolumn}
\usepackage{bm}
\usepackage{color}
\usepackage{epstopdf}
\usepackage{epsfig}
\usepackage[colorlinks=true, citecolor=blue, anchorcolor=green]{hyperref}
\usepackage[normalem]{ulem}

\makeatletter
\begin{document}
\preprint{APS/123-QED}
\title{Spatiotemporal imaging of microwave magnetic fields via magnonic coherent splitting}

\author{C.K.~Wei}\email{Contributed equally to this work}
\affiliation{State Key Laboratory of Infrared Physics, Shanghai Institute of Technical Physics, Chinese Academy of Sciences, Shanghai 200083, China}

\author{Z.J.~Chen}\email{Contributed equally to this work}
\affiliation{School of Physical Science and Technology, ShanghaiTech University, Shanghai 201210, China}

\author{J.T.~Song}
\affiliation{School of Physical Science and Technology, ShanghaiTech University, Shanghai 201210, China}

\author{S.H.~Ma}
\affiliation{School of Physical Science and Technology, ShanghaiTech University, Shanghai 201210, China}

\author{W.H.~Liu}
\affiliation{State Key Laboratory of Infrared Physics, Shanghai Institute of Technical Physics, Chinese Academy of Sciences, Shanghai 200083, China}

\author{J.H.~Wu}
\affiliation{School of Physical Science and Technology, ShanghaiTech University, Shanghai 201210, China}

\author{Z.W.~Huang}
\affiliation{School of Physical Science and Technology, ShanghaiTech University, Shanghai 201210, China}

\author{Jinwei~Rao}\email{raojw@sdu.edu.cn}
\affiliation{School of Physics, Shandong University, Jinan 250100,  China}

\author{Wei~Lu}\email{luwei@mail.sitp.ac.cn;}
\affiliation{State Key Laboratory of Infrared Physics, Shanghai Institute of Technical Physics, Chinese Academy of Sciences, Shanghai 200083, China}
\affiliation{School of Physical Science and Technology, ShanghaiTech University, Shanghai 201210, China}

\author{Bimu~Yao}\email{yaobimu@mail.sitp.ac.cn;}
\affiliation{State Key Laboratory of Infrared Physics, Shanghai Institute of Technical Physics, Chinese Academy of Sciences, Shanghai 200083, China}
\affiliation{School of Physical Science and Technology, ShanghaiTech University, Shanghai 201210, China}

\begin{abstract}

Spatiotemporal microwave magnetic-field imaging reveals current flow in high-frequency circuits and nonequilibrium spin dynamics\cite{ogier2026imaging,chen2021focusing,rao2025time,zhang2014strongly,joseph2026dynamic,woltersdorf2007microwave, bertelli2020magnetic,jarmola2012temperature}, yet probes rarely combine calibrated spectral readout, optics-free operation and transient mapping at room temperature\cite{bell1961optically}. Here ferrimagnetic order in yttrium iron garnet supports coherent coupling from a pump-induced magnon mode\cite{tabuchi2014hybridizing,rao2023unveiling}, converting target-field amplitude into a spectral splitting with all-microwave readout. Sampling the calibrated splitting over position and delay reconstructs spatiotemporal imaging of magnetic fields. Continuous-wave measurement reaches a sensitivity of $58~\text{pT}/\sqrt{\text{Hz}}$ and recovers phases across various powers. Combined with time-resolved frequency-comb spectroscopy\cite{rao2025time}, the method reconstructs transient fields with a 130-ns response time. Coplanar-waveguide imaging validates the magnetic selectivity of the mode-splitting readout, where measured maps agree with simulated magnetic-field distribution. In a microwave amplifier, our reconstruction resolves nonuniform switching dynamics and detects downstream field suppression from an open-contact fault. Magnonic coherent splitting  provides a scalable route for optics-free imaging of spatiotemporal field evolution in functional devices.

\end{abstract}

\maketitle

Microwave magnetic fields sit at the interface between high-frequency electronics and coherent spin physics: they are generated by radio-frequency (RF) currents and in turn drive spin excitations in magnetic and quantum systems\cite{tabuchi2014hybridizing,zhang2014strongly,PhysRevLett.111.127003,li2019nutation,huang2024high}. Their spatial structure encodes the pathways, interference patterns and local discontinuities of high-frequency currents\cite{chen2021focusing,ogier2026imaging,liu2025reconfigurable}, whereas their temporal evolution records the transient switching, fast propagation, and rapid relaxation of spin dynamic systems\cite{rao2025time,zhang2014strongly,woltersdorf2007microwave, bertelli2020magnetic,jarmola2012temperature,joseph2026dynamic}. Direct spatiotemporal access to magnetic fields is therefore sought for non-destructive diagnosis of operating RF circuits and real-space studies of spin dynamic systems\cite{wang2015high,zhou2021magnon,bertelli2020magnetic}, particularly informative for near fields where the dominant magnetic component\cite{ott2011electromagnetic} cannot generally be inferred from conventional electric-field measurements alone.

Existing approaches access different parts of this spatiotemporal picture. Miniaturized loops provide optics-free detection through direct measurement of the induced-voltage, but this readout is constrained by thermal and electronic noise\cite{li2013effect}. Coherent spin sensors provide a complementary route by encoding magnetic fields into spectral splitting\cite{balasubramanian2008nanoscale,wang2015high,eisenach2021cavity}. Atomic-vapor cells \cite{horsley2016frequency,sun2018tunable} and nitrogen-vacancy (NV) centers \cite{loretz2013radio,horsley2018microwave,tang2023quantum,herb2025quantum,zopes2019reconstruction,xu2016wavelet,cooper2014time} have enabled broadband microwave detection\cite{carmiggelt2023broadband}, on-chip electromagnetic-interference testing \cite{ma2023surface,yin2022quantum} and local probes of GHz spin dynamics\cite{rable2023off}. Their reliance on optical preparation and readout\cite{bell1961optically}, however, adds system complexity, while limited optical contrast and photon collection require substantial averaging. For wide-field imaging, the accessible area is constrained by the sensing medium, optical coverage and material uniformity \cite{shao2016wide,horsley2018microwave,zhang2026high}. These limitations call for a room-temperature, all-microwave sensor that converts local magnetic fields into calibrated spectral splittings for spatiotemporal reconstruction.

Coherent coupling between a pump-induced magnon mode (PIM)\cite{rao2023unveiling} and a Walker mode (WM) \cite{walker1957magnetostatic} in yttrium iron garnet (YIG) provides a direct solution. YIG’s room-temperature ferrimagnetic order provides a microwave-addressable collective magnetization without optical preparation\cite{tabuchi2014hybridizing}. The target microwave magnetic field excites the PIM, which hybridizes with the WM and produces a spectral splitting that can be read in transmission without optical detection. The PIM-WM coupling strength $g$ scales as the square root of the target field amplitude, providing a calibrated field-to-splitting transduction with a steep slope at low field amplitudes\cite{rao2023unveiling,wang2024enhancement}. Approximately $10^{19}$ spins and low damping coefficient $10^{-5}$ in the insulating medium make the splitting sensitive and resolvable, while magnetic tunability matches the resonance to the target frequency\cite{tabuchi2014hybridizing, zhang2014strongly, chumak2015magnon, PhysRevLett.111.127003,zhang2019experimental,yao2019coherent}. By naturally integrating time-resolved frequency-comb spectroscopy (trFCS) \cite{rao2025time}, which resolves spectral splittings at successive delays, we can combine spatial and delay scans to reconstruct $B(\mathbf{r},\tau)$ with simultaneous spatial and temporal resolution.

Here we demonstrate room-temperature, optics-free reconstruction of microwave magnetic fields across position and time in an open geometry. The same calibrated PIM–WM splitting serves as the field observable at every position and delay, converting raster and delay scans into $B(\mathbf{r},\tau)$. We establish field transduction under continuous-wave operation, reaching a sensitivity of $58~\text{pT}/\sqrt{\text{Hz}}$. Combination with trFCS enables transient fields to be reconstructed with a response time of 130 ns. Above an open-ended coplanar waveguide (CPW), our reconstruction matches the simulated magnetic-field pattern rather than its electric-field counterpart and follows the independently recorded temporal envelope. Applied to an operating commercial RF amplifier, our method reveals spatially nonuniform near-field evolution and temporal latencies during switching, and demonstrates the diagnosis of an open-contact fault. With an image extent determined by the mechanical scan range, our approach provides a scalable route to non-contact spatiotemporal characterization of operating microwave devices.

\begin{figure}[htbp]
\centering
\epsfig{file=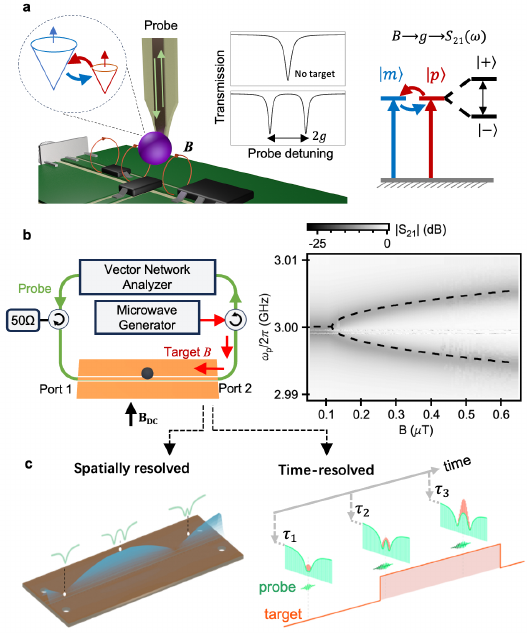,width=8.4cm}
\caption{\textbf{Field-to-splitting transduction for spatiotemporal microwave magnetic-field imaging.}
\textbf{a}, A target microwave magnetic field $B$ excites a PIM $\lvert p\rangle$ in a 1-mm-diameter YIG sphere. The PIM couples coherently to a Walker mode (WM, $\lvert m\rangle$) and produces the hybrid modes $|-\rangle$ and $|+\rangle$ whose on-resonance separation approaches $2g$, where $g$ is the PIM–WM coupling strength. It allows the target-field amplitude $B$ to be quantitatively reconstructed from spectral fitting.
\textbf{b}, Calibration of the field-to-splitting relation. A DC bias field $B_{\mathrm{DC}}$ sets the WM resonance, the target field excites the PIM, and a weak probe reads out the transmission amplitude $|S_{21}|$. Increasing $B$ increases the hybrid-mode separation as we measured in the mapping. The black dashed curves show $(\omega_d\pm g)/2\pi$, with $g$ following the relation $g(B)\propto\sqrt{B-B_0}$, where $B_0=115.7~\mathrm{nT}$ is the fitted threshold of splitting.
\textbf{c}, Spatial and time-resolved imaging protocols. Raster scanning yields $B(\mathbf{r})$, whereas delay-scanned trFCS reconstructs $B(\mathbf{r},\tau)$ from a periodically modulated target field.}
\label{fig:schematic}
\end{figure}

\section{Results}
\subsection{Field-to-splitting transduction}

In the PIM–WM system, the target microwave field excites the PIM, whose coherent coupling $g$ to the Walker mode produces a transmission doublet with an on-resonance separation approaching $2g$. Since $g$ depends on the target-field amplitude $B$, this splitting provides the common local observable used throughout this work (Figure~\ref{fig:schematic}(a)). A DC bias field $B_{\mathrm{DC}}$ tunes the Walker mode of a 1-mm-diameter YIG sphere to the target frequency $\omega_d$, while the transverse target-field component excites the PIM and a weak microwave probe reads out the hybrid modes.
We extract $g$ by fitting the weak-probe transmission spectra to a coupled-mode model (further details in Supplementary Note 1). To calibrate $g$ against $B$, we place the sphere on a CPW driven at $\omega_d/2\pi=3~\mathrm{GHz}$ and record the corresponding spectra $|S_{21}|$ by successively varying $B$ (Fig.~\ref{fig:schematic}(b)). We extract the coupling strength $g$ and plot $(\omega_d\pm g)/2\pi$ as dashed lines in Fig.~\ref{fig:schematic} (b). The fitted coupling follows $g(B)= 2\pi A\sqrt{B-B_0}$, where A is field-to-splitting transduction coefficient and $B_0$ is the fitted threshold for coherent splitting\cite{rao2023unveiling,Qu_2023,Qu_2025,jnpb-2mxx}. This relation defines the common conversion from measured $g$ to $B$ used in all subsequent spatial and time-resolved reconstructions.

The field-to-splitting relation enables the YIG sphere to operate as a scanning microwave magnetic-field probe (Fig.~\ref{fig:schematic}(c)). Raster scanning and spectral fitting yield $g(\mathbf{r})$, which is converted into the field map $B(\mathbf{r})$ (Supplementary Note 3). For time-resolved imaging, we combine the spatial scan with delay-scanned trFCS \cite{rao2025time} (Supplementary Note~4). Periodic probe pulses generate a microwave frequency comb whose tooth amplitudes are used to reconstruct the transmission spectrum (green lines in Fig.~\ref{fig:schematic}(c)), while the delay $\tau$ relative to the target waveform selects the sampled time. Repeating the measurement at each position and delay yields a reconstructed field map $B(\mathbf{r},\tau)$.

\begin{figure}[htbp]
\centering
\epsfig{file=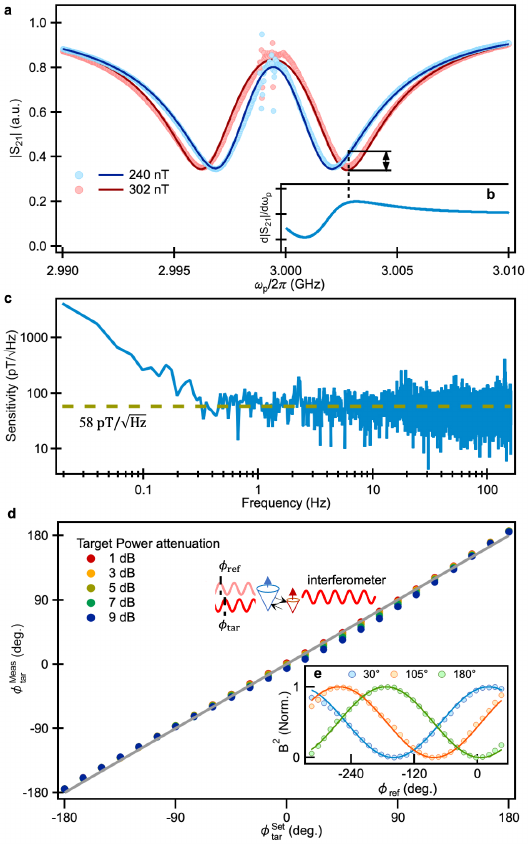,width=0.9\columnwidth}
\caption{\textbf{Continuous-wave sensitivity and interferometric phase readout of the microwave magnetic-field sensor.} 
\textbf{a}, Weak-probe transmission spectra at target-field amplitudes $B=240~\mathrm{nT}$ and $302~\mathrm{nT}$, showing the field-dependent separation of the hybrid modes. Symbols and solid lines represent measurements and fits. \textbf{b}, Frequency derivative of the high-frequency hybrid branch $|+\rangle$ at $B=240~\mathrm{nT}$. The fixed readout frequency $f_{\mathrm{meas}}$ is selected at the maximum of $\mathrm{d}|S_{21}|/\mathrm{d}\omega_p$.
\textbf{c}, Magnetic-field sensitivity at the operating field $B=240~\mathrm{nT}$ as a function of Fourier frequency. The dashed line marks the sensitivity in the white-noise region, $58~\mathrm{pT}/\sqrt{\mathrm{Hz}}$.
\textbf{d}, Phase readout based on microwave interference. The measured target phase $\phi_{\rm tar}^{\mathrm{meas}}$ is plotted against the set phase $\phi_{\rm tar}^{\mathrm{set}}$ under different power attenuations. The grey line denotes the ideal response ($\phi_{\rm tar}^{\mathrm{meas}}=\phi_{\rm tar}^{\mathrm{set}}$). 
\textbf{e}, Extracted normalized squared field amplitude $B^2$ as a function of reference phase $\phi_{\rm ref}$ for three different set phases, $30^\circ$, $105^\circ$ and $180^\circ$.}
\label{fig:magphase}
\end{figure}

\begin{figure*}[htbp]
\centering
\epsfig{file=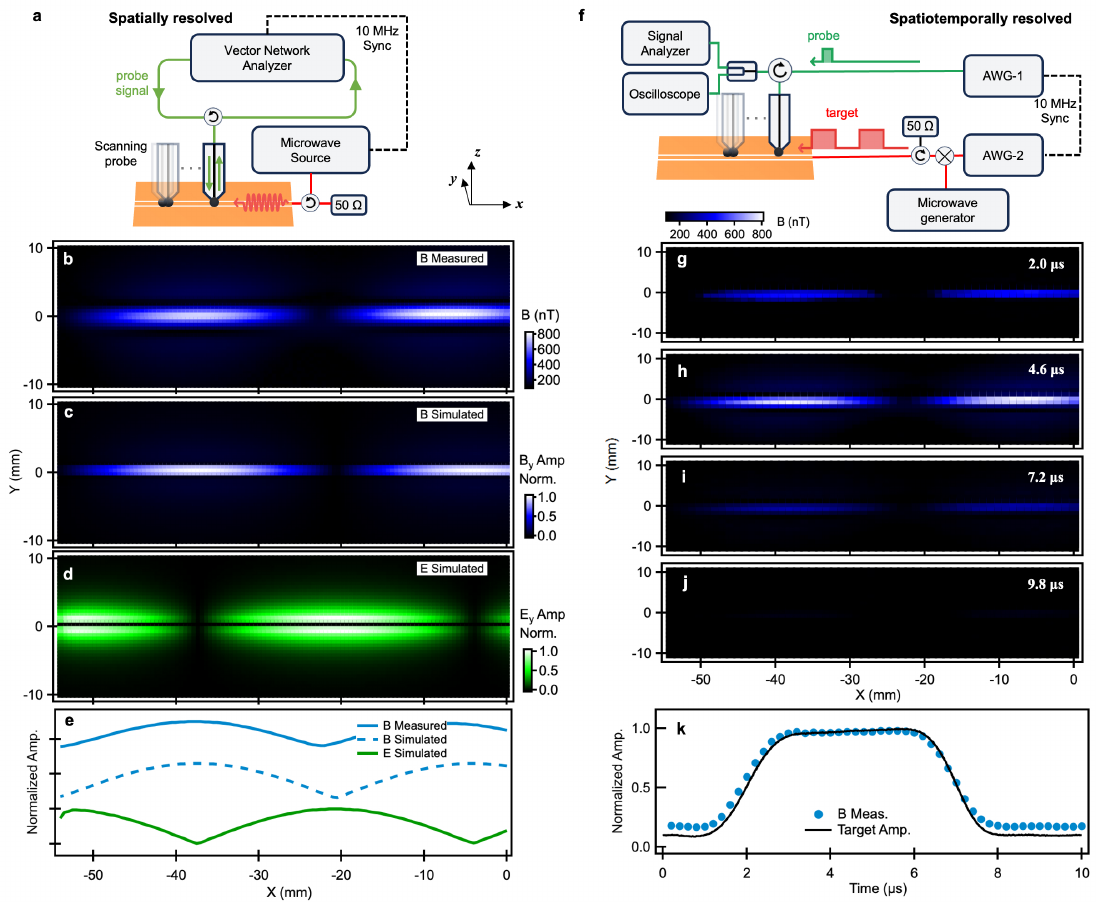,width=17.5cm}
\caption{\textbf{Steady-state and time-resolved imaging of microwave magnetic fields above a CPW.} 
\textbf{a}, Steady-state scanning configuration. A PIM-based probe, biased by a permanent magnet along $z$ to set its WM resonance, is scanned above an open-ended CPW driven by a continuous-wave tone near 3 GHz, producing a standing-wave target field.
\textbf{b-d}, Field maps at a probe height of $2~\mathrm{mm}$: measured magnetic-field amplitude (\textbf{b}), simulated magnetic-field amplitude $|B_y|$ (\textbf{c}) and simulated electric-field amplitude $|E_y|$ (\textbf{d}). 
\textbf{e}, Line cuts from panels \textbf{b-d} at $0.5~\mathrm{mm}$ from the CPW centre line. The measured profile follows the simulated magnetic-field distribution and is spatially offset from the electric-field profile.
\textbf{f}, Time-resolved imaging configuration using trFCS with synchronized target-field and probe pulse sequences from arbitrary waveform generators (AWGs). The probe pulse spectrum is acquired with a spectrum analyser, and an oscilloscope defines the temporal origin.
\textbf{g-j}, Typical field maps reconstructed at delay times of $2$, $4.6$, $7.2$, and $9.8~\mu\mathrm{s}$, respectively, showing the turn-on, steady state and decay of the standing-wave envelope.
\textbf{k}, Normalized field amplitude at $(x,y)$=(-37 mm,0 mm) compared with the independently recorded normalized microwave-drive envelope, showing that the measured rise/fall response follows the drive envelope.}
\label{fig:imaging}
\end{figure*}

We next quantify the sensitivity of this calibrated spectral readout at an operating field $B=240~\mathrm{nT}$. Increasing $B$ from $240$ to $302~\mathrm{nT}$ modifies the PIM–WM coupling strength and consequently shifts the eigenfrequencies as shown in Fig.~\ref{fig:magphase}(a). The shift of the eigenfrequencies of the coupled PIM–WM system manifests as an amplitude change at the readout frequency $f_{\mathrm{meas}}$, which is set at the maximum of $\mathrm{d}|S_{21}|/\mathrm{d}\omega_p$ of the high-frequency hybrid branch to maximize the conversion of small spectral shifts into changes in $|S_{21}|$ (marked by the black dashed line in Fig.~\ref{fig:magphase}(b)).  The Allan deviation of the continuous readout\cite{wolf2015subpicotesla,Riley2008FrequencyStability,stuerner2021} follows a $T_{\rm avg}^{-1/2}$ dependence over $T_{\rm avg}=3.1$--$431.3~\mathrm{ms}$, where $T_{\rm avg}$ denotes the averaging time. This behaviour is consistent with white-noise-limited averaging, and the corresponding bandwidth range is $1.16$--$160~\mathrm{Hz}$. We divide the noise amplitude spectral density (ASD) by the calibrated transduction slope $\left|\mathrm{d}|S_{21}|/\mathrm{d}B\right|$ to obtain a room-temperature magnetic-field sensitivity of $58~\text{pT}/\sqrt{\text{Hz}}$ in Fig.~\ref{fig:magphase}(c) (see Supplementary Note~2 for details).

To determine the target-field phase $\phi_{\mathrm{tar}}$ with the same spectral readout, we add a coherent reference field with a tunable phase $\phi_{\mathrm{ref}}$. At the YIG sphere, interference between the target and reference fields gives the total squared field amplitude a $\cos(\phi_{\mathrm{ref}}-\phi_{\mathrm{tar}})$ dependence, and the resulting amplitude variation is read through the fitted coupling $g$. Sweeping $\phi_{\mathrm{ref}}$ therefore produces a sinusoidal variation in the normalized squared field amplitude $B^2$, from which $\phi_{\mathrm{tar}}$ is obtained by fitting the initial phase as shown in Fig.~\ref{fig:magphase}(e) typically for $30^\circ$, $105^\circ$ and $180^\circ$. Across the tested target-field attenuations, the recovered phase $\phi_{\mathrm{tar}}^{\mathrm{meas}}$ follows the set phase $\phi_{\mathrm{tar}}^{\mathrm{set}}$ with unit slope and zero intercept displayed in grey line in Fig.~\ref{fig:magphase}(d), demonstrating accurate phase retrieval as the target-field amplitude varies (details in Supplementary Note 2).

\subsection{Spatiotemporal field reconstruction}

We next test the spatial imaging and magnetic selectivity of the calibrated reconstruction using an open-ended CPW with a well-defined standing-wave near field, as sketched in Fig.~\ref{fig:imaging}(a). For imaging, the YIG sphere is attached to the end of the scanning probe and positioned above the CPW. Its simple field geometry, with a predominantly transverse microwave magnetic field along $y$ above the signal line, provides a convenient comparison between experiment and simulation. A continuous-wave tone near $3~\mathrm{GHz}$ is launched into the CPW and reflected at the open termination due to impedance mismatch, producing spatially offset electric- and magnetic-field nodes\cite{xu2021controlling}. We scan the probe in the $xy$ plane at a fixed height of $2~\mathrm{mm}$ above the CPW. To maintain a consistent resonant sensing condition, the drive frequency is locally adjusted to track the WM resonance at each scan position, with a total variation across the scan below $0.2\%$ of the signal frequency.

Fig.~\ref{fig:imaging}(b) shows the measured $xy$-plane map. It closely matches the simulated magnetic-field $\lvert B_y\rvert$ distribution in Fig.~\ref{fig:imaging}(c), but is spatially offset from the simulated electric-field $\lvert E_y\rvert$ distribution in Fig.~\ref{fig:imaging}(d). We take horizontal cuts through Figs.~\ref{fig:imaging}(b--d) along a line offset by $0.5~\mathrm{mm}$ from the centre to retain characteristic features of both field distributions, and compare the resulting profiles in Fig.~\ref{fig:imaging}(e). The measured profile agrees with the simulated $\lvert B_y\rvert$ profile but deviates from the simulated $\lvert E_y\rvert$ profile, confirming that the probe responds to the magnetic field rather than the electric field. A control scan using the microwave probe alone without the YIG sphere instead matches the simulated $\lvert E\rvert$ distribution rather than the corresponding $\lvert B_y\rvert$ distribution, excluding direct magnetic-field pickup by the microwave probe. Additional scans in the orthogonal $xz$ and $yz$ planes demonstrate near-field mapping in three-dimensional spatial planes (Supplementary Note~3).

\begin{figure*}[htbp]
\centering
\epsfig{file=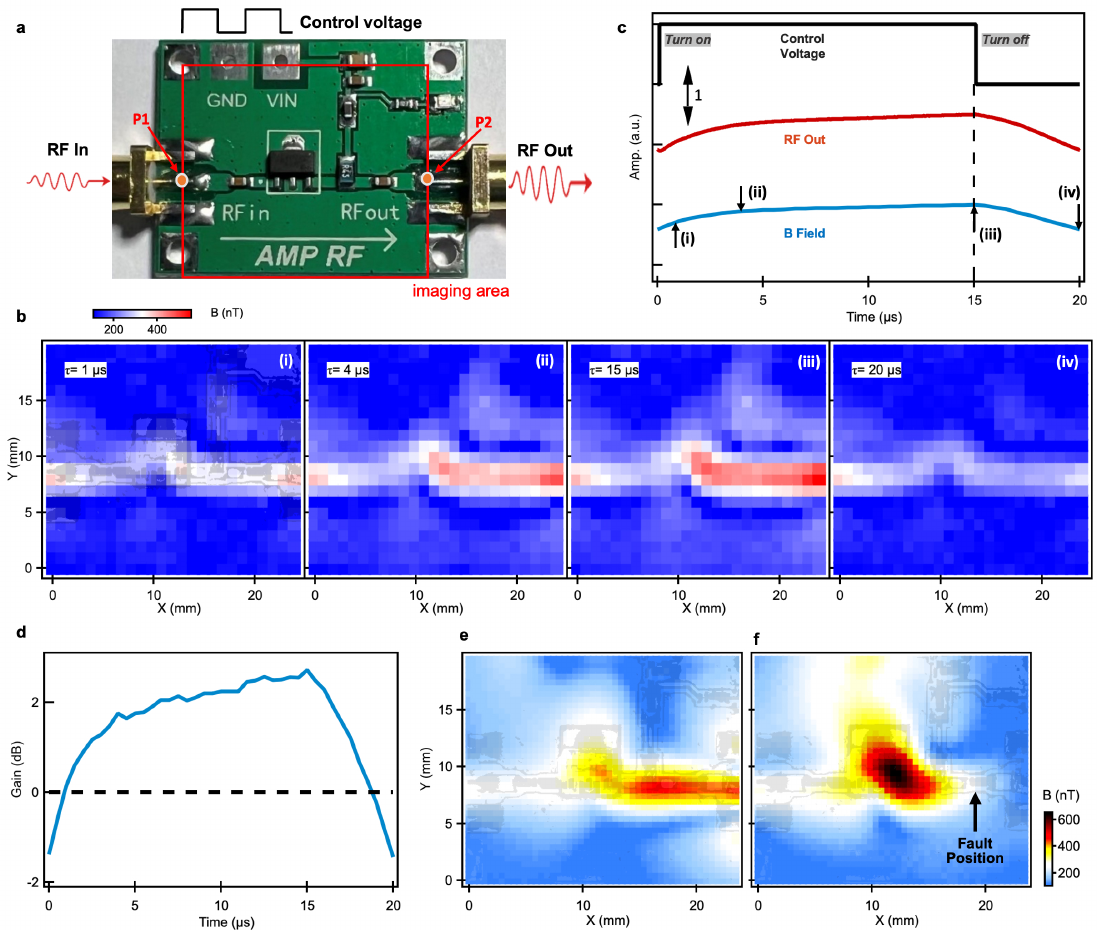,width=17cm}
\caption{\textbf{Spatiotemporal microwave magnetic-field imaging of an operating RF amplifier.} 
\textbf{a}, Commercial RF amplifier and scanned region. A continuous RF tone is applied at the input, while a pulsed control voltage between input-bias terminal (VIN) and ground (GND) switches the amplifier gain. $\mathrm{P_1}$ and $\mathrm{P_2}$ denote probe positions near the input and output sides, respectively.
\textbf{b}, Time-resolved magnetic-field maps measured at delays of 1, 4, 15 and $20~\mu\mathrm{s}$.
\textbf{c}, Normalized control-voltage, RF output-voltage and target magnetic-field traces measured at $\mathrm{P_2}$. The target magnetic field follows the RF output envelope and exhibits 10--90\% rise and 90--10\% fall times of $9.5~\mu\mathrm{s}$ and $3.5~\mu\mathrm{s}$, respectively.
\textbf{d}, Time-dependent output-to-input gain ratio between $\mathrm{P_2}$ and $\mathrm{P_1}$. 
\textbf{e-f}, Measured steady-state magnetic-field maps of a normal amplifier (\textbf{e}) and a second amplifier after one terminal of its output DC-blocking capacitor was lifted (\textbf{f}). The black arrow marks the fault position. A semitransparent device photograph is overlaid on panels \textbf{b(i)}, \textbf{e}, and \textbf{f}.}
\label{fig:tramp}
\end{figure*}

To test the temporal imaging capability of our method, we adopt the trFCS\cite{rao2025time} technique to detect a periodically modulated target field at successive delays $\tau$, as illustrated in Fig.~\ref{fig:imaging}(f). We repeat the measurement at different target–probe delays and stroboscopically reconstruct $B(\mathbf{r},\tau)$. The maps are sampled at 200-ns intervals, whereas the magnon damping rate sets a 130-ns sensor response time\cite{xu2024strong}. Representative frames in Figs.~\ref{fig:imaging}(g)–(j) visualize the build-up, steady state and decay of the standing-wave magnetic-field in the open-ended CPW, evolving from a signal level consistent with the noise floor to the established standing-wave pattern and back after the drive is switched off (the complete spatiotemporal imaging is provided in supplementary video 1). To quantify the temporal response, we compare the field amplitude at $(x,y)$=(-37 mm,0 mm) within the imaging region with the independently recorded microwave-drive envelope in Fig.~\ref{fig:imaging}(k). The measured field closely follows the drive waveform, validating the temporal fidelity of the reconstruction.

\subsection{Imaging RF-amplifier operation}

Having demonstrated the spatiotemporal imaging on the CPW, we next apply the same method to reconstruct the magnetic-field amplitude $B(\mathbf{r},\tau)$ of a commercial RF-amplifier operating under pulsed gain control (see Supplementary Note~5), as sketched in Fig.~\ref{fig:tramp}(a). Quantitative prediction of the transient near field of an RF amplifier is not yet feasible, as it requires detailed geometric and bias-dependent device models. The amplifier therefore serves as a practical test case for spatiotemporal imaging of magnetic fields under active switching conditions.

We input a continuous RF tone into the amplifier while a pulsed bias voltage is used to switch its gain. Repeating the delay-scanned measurement at each probe position reconstructs the effective magnetic-field amplitude $B(\mathbf{r},\tau)$ over a $0$--$20~\mu\text{s}$ window referenced to the control-waveform trigger, as shown in the typical frames in Fig.~\ref{fig:tramp}(b), complete spatiotemporal imaging is provided in supplementary video 2. Displayed on a common colour scale, these maps show the field emerging from the measurement noise level and developing a localized hotspot over the amplifier chip and output region before decaying after the bias is turned off. The position of the field maximum on the output side remains within $1~\mathrm{mm}$ of the amplifier output port, while its amplitude increases by $32.8\%$ over the first $1.5~\mu\mathrm{s}$ (Supplementary Note~5). These time-resolved maps reveal the spatially nonuniform build-up and decay of the microwave magnetic field during amplifier switching.

Our time-resolved approach captures the amplifier's non-equilibrium working conditions. Figure~\ref{fig:tramp}(c) compares the target magnetic-field amplitude (blue line) at the output-side position $\mathrm{P_2}$ marked in Fig.~\ref{fig:tramp}(a) with the applied control waveform (black line) and the independently recorded RF output-voltage envelope (red line). After normalization, the measured local magnetic-field trace follows the RF output envelope, validating the temporal fidelity of the near-field readout. The measured local magnetic field exhibits 10--90\% rise and 90--10\% fall times\cite{hu2024environmental} of $9.5~\mu\mathrm{s}$ and $3.5~\mu\mathrm{s}$, respectively (Supplementary Note~5). From the local field amplitudes at the output-side $\mathrm{P_2}$ and input-side $\mathrm{P_1}$ (indicated in Fig.~\ref{fig:tramp}(a)), we calculate the time-dependent gain ratio $G(\tau)=10{\rm log}_{10}[B^2({\rm P_{2}},\tau)/B^2({\rm P_{1}},\tau)]$ (Fig.~\ref{fig:tramp}(d)). These values characterize the transient response of the driven amplifier, which goes beyond steady-state characterization. To demonstrate fault diagnosis, we compare the magnetic field distributions of a normal amplifier and a second amplifier after one terminal of the DC-blocking capacitor on the output port was lifted (indicated by the black arrow in Fig.~\ref{fig:tramp}(f)). Compared with the normal-state map, the fault-state map shows a strong suppression of the field downstream of the capacitor, consistent with an interruption of the RF output path (Fig.~\ref{fig:tramp}(e) and (f)).

\textsl{Conclusions}.---In summary, we establish coherent PIM–WM splitting as a calibrated spectral observable for all-microwave reconstruction of microwave magnetic fields across position and time. Under continuous-wave operation at room temperature, the readout reaches a sensitivity of $58~\text{pT}/\sqrt{\text{Hz}}$. In separate delay-resolved measurements, it reconstructs repeatable transients with a response time of 130 ns set by the magnon linewidth. The CPW comparison identifies the reconstructed spatial pattern of the magnetic field rather than electric field. The reconstructed field’s temporal evolution matches an independently recorded drive waveform. These comparisons support the magnetic selectivity and spatiotemporal fidelity of the reconstruction. Applied to an operating commercial RF amplifier, the same reconstruction resolves spatially nonuniform switching fields, relates local field evolution to the output waveform and time-dependent gain, and reveals downstream field suppression caused by a controlled open-contact fault. With the field of view set by the scan range rather than the sensing-medium dimensions or optical coverage, the approach offers a scalable route for advanced RF-circuit diagnostics, high-frequency spintronic studies and non-destructive, contact-free testing.

\section{Data availability}
The data that support the findings of this study are available from corresponding authors upon reasonable request.

\bibliography{ref}

\section{Acknowledgements}
This work has been supported by the Strategic Priority Research Program of the Chinese Academy of Sciences (Grant No. XDB0580000), STCSM (Grant No. 23JC1404100), the National Natural Science Foundation of China (Grant Nos. 12227901, U23A6002, 12522405, 12474120), the Shandong Provincial Natural Science Foundation, China (Grant No. ZR2024YQ001), and the Taishan Young Scholar Program of Shandong Province (Grant No. tsqn202507065).

\section{Author contributions}

B.M.Y. and J.W.R. conceived this study,  B.M.Y., C.K.W., Z.J.C. and J.W.R. designed the experimental set-up. C.K.W., Z.J.C., J.T.S., S.H.M. and B.M.Y. performed the measurements and data analysis. C.K.W. and J.T.S. contributed to the theoretical analysis and validation. J.H.W. contributed to the theoretical derivation relating power to magnetic-field strength. C.K.W., Z.J.C., J.T.S., S.H.M., W.H.L., J.H.W., Z.W.H., J.W.R., W.L. and B.M.Y. together contributed to the writing of the paper. W.L. provided conceptual advice and revised the manuscript. B.M.Y. supervised this work.

\section{Competing interests}
The authors declare no competing interests.

\section{Correspondence}
Correspondence and requests for materials should be addressed to Jinwei Rao, Bimu Yao and Wei Lu. 
\end{document}